\documentstyle[prl,aps,multicol,epsfig]{revtex}

\newcommand{\beq}{\begin{equation}}
\newcommand{\eeq}{\end{equation}}
\newcommand{\beqn}{\begin{eqnarray}}
\newcommand{\eeqn}{\end{eqnarray}}

\begin{document}


\title{Softening and melting of a vortex lattice in presence of point disorder.} 
\author{Denis {\sc Feinberg}} 
\address{Laboratoire d'Etudes des Propri\'et\'es Electroniques des Solides, Centre National de la Recherche Scientifique,\\ {\it
BP166  38042 Grenoble Cedex} France} \date{\today}

\maketitle 

\begin{abstract}
A phenomenological model is proposed for melting of a vortex lattice, based on screening of the elastic shear modulus by mobile or partially pinned dislocations. A first-order softening line is found and ends at a critical point
 beyond which the lattice crosses over to an hexatic vortex solid. The consequences of softening on vortex dynamics are explored, as fingerprints of plastic dynamics:
 a reentrance of single vortex behaviour, for both depinning and collective creep, occurs as the field increases, with non-monotonous creep exponents. This general scenario is supported by recent experiments in high-$T_c$ materials and suggests that for a 3D vortex lattice at low temperature the field induces a continuous order-disorder transition towards a glassy phase.
\bigskip


\bigskip

\end{abstract}
\begin{multicols}{2}

The interplay between thermal fluctuations and quenched material disorder controls the phase diagram of vortex matter in High Temperature Superconductors \cite{bible}. In clean crystals, a first-order melting line exists but ends at a critical point \cite{melt,safar}, while in strong fields or 
 for strong enough material disorder, the vortex lattice (VL) melts continuously, possibly through a vortex glass (VG) transition \cite {vg}. It has been recently proposed that a solid-solid transition also occurs with increasing field, between a quasi-ordered ``Bragg glass''
 and a disordered vortex solid \cite {Giam,Kierf,EN}. This is supported by irreversible magnetization measurements showing an increase of the effective critical current (``second peak'') \cite{Peak,Delig} and by neutron diffraction experiments \cite{Forgan} .
 Such a transition is attributed to the proliferation of disorder-induced topological defects. However, contrarily to the pure melting transition, its first-order character might be questioned, since those defects are also {\it pinned} by
 point disorder and therefore will not dramatically screen the elastic shear modulus, contrarily to a pure melting scenario.
 In what follows we propose a simple model of plastic behaviour, based on short range fluctuations. It produces a first-order transition with a critical point, by capturing the two effects of point disorder, e.g. promoting {\it and} pinning dislocations. From the calculated renormalized shear modulus, vortex dynamics can be discussed.\\
\indent
Let us consider the softening of the shear modulus $C_{66}=\frac{\phi_0B}{(8\pi \lambda)^2}$($\phi_0=\frac{hc}{2e}$, $\lambda$ is the London length with field parallel to the c-axis for HTSC). For simplicity the $T$- dependence of $C_{66}$ is neglected. As shown by Brandt \cite{Brandt} and Marchetti and Nelson \cite{MN}, softening is due to motion of edge dislocation (DL) segments normally to the field. 
These are part of DL loops with screw components normal to the field. Their natural unit length is $L_0 \sim (\frac{\epsilon_0}{C_{66}})^{1/2} \sim \frac{a_0}{\gamma}$, the elastic intervortex ``interaction length'' \cite{bible}. Here $a_0$ and $\gamma$ are the vortex spacing and the anisotropy coefficient respectively and $\epsilon_0$ is the vortex line stiffness.
 Rather than a detailed description of these loops, let us forget their line character and attempt at a mean-field picture, valid in principle for high DL densities
. First, the edge and screw core energies being roughly equivalent, on can simply calculate the energy and average density of screw DL cores. 
Those can be considered as ``kinks'' where a vortex line passes from a valley to another of a periodic ``egg-carton'' potential representing the reference perfect VL on top of which DL's are created. 
This extended potential should replace the parabolic ``cage'' potential, valid only for the elastic analysis of the fluctuations of one vortex line in the mean field of the others \cite{Kierf,EN}. In what follows, we shall more simply  obtain the kink density from a line's fluctuations in a {\it renormalized} cage potential.
The kink energy is $E_K \sim \alpha E_{el}$ where $E_{el} \sim 4c^2_L C_{66}a^2_0L_0$ is the typical elastic cohesive energy ($\alpha > 1$ is a numerical constant and $c_L \sim 0.15$ the Lindemann constant). On the kink length $\sim L_0$, a vortex line experiences gaussian lateral fluctuations $<u^2> \sim \frac{L_0}{\epsilon_0}(T+\Delta)$ where $\Delta \sim U_0 (\frac{L_0}{L_c})^{2\zeta-1}$ is the elastic  pinning energy on the length $L_0$ \cite{bible,EN}. Here $L_c$ and $U_0$ are the line's Larkin length and pinning energy respectively. We shall consider here only independently pinned vortices. i.e. the single vortex (SV) regime for which $L_0 > L_c$, then $\zeta \sim 3/5$ \cite{bible}. 
This allows to write the DL density as $n_d \sim \frac{1}{a_0^2 L_0}\exp(-\frac{\alpha E_{el}}{T+\Delta}) \sim \frac{1}{a_0^2 L_1}$ where $L_1$ is a longitudinal ``domain size'', and $R_1 \sim a_0 \exp(\frac{\alpha E_{el}}{T+\Delta})$ is a transverse positional correlation length. The picture here is that of an {\it hexatic} VL since DL's preserve orientational order.\\
\indent
 Let us now turn to the screening phenomenon. If edge DL's were free to move, they would perfectly relax shear stresses and renormalize $C_{66}$ to zero \cite{MN}. However, they experience two types of potentials. The first one is due to the VL periodicity and builds the Peierls-Nabarro barriers, of order $\nu E_{el}$ ($\nu$ of order 1). The second one is the disorder pinning potential, which in the SV regime acts on DL cores of length $L_0$, thus is {\it also} of order $\Delta$ and has a correlation length $\Xi_0 \sim a_0$.  
In the bundle or 3D regime ($L_c > L_0$) the situation is more complicated : the part of the DL pinning energy coming from elastic displacements of neighbouring vortices builds a random potential which has {\it a priori} a correlation length $\Xi_0 > a_0$. 
 For a dense collection of DL's, DL interactions will lead to collective pinning effects, e.g. new pinning lengths will appear. 
Such`` plastic'' correlation lengths should be considered in a refined analysis.\\
\indent
In the above picture, the energy and length scales are the same as in the elastic analysis. 
Nevertheless, the self-consistent screening mechanism leads to qualitatively different results for the melting line. 
To analyse the softening of the vortex solid by proliferating DL's, let us define a renormalized shear modulus $C_{66}^R = x C_{66} (x<1)$. Then all relevant ``elastic'' quantities must be accordingly modified as $L^R_0=x^{-1/2}L_0 > L_0$, $E^R_{el}=E_{el} x^{1/2} < E_{el}$ and $\Delta^R=\Delta x^{-1/10} > \Delta$. 
Softening therefore increases the density of DL's, which in absence of disorder drives the first-order melting transition. To close the self-consistency loop one must express the softening of $C_{66}$ by moving DL's.
 This can be done similarly to Ref. \cite{MN}, though in a simplified way. Indeed, vortex line displacements are planar and one can write a simplified screening equation for each ``layer'' of vortex matter of thickness $L^R_0$, e.g. $\frac{1}{C^R_{66}} = \frac{1}{C_{66}} + \chi$
where $\chi = \frac{1}{T}N_d  L^R_0<u^2>_d$ is a ``plastic'' susceptibility due to motion of $N_d \sim \exp(\frac{\alpha E_{el}}{T+\Delta})$ edge DL segments of length $L^R_0$. The correlation function $<u_d^2>$ of DL displacements can be evaluated,
 given the potential well (Peierls-Nabarro plus disorder) in which they sit. One can write 
 $<u^2_d> \sim a^2_0 y f(y)$ with $y=\frac{T}{\nu E_{el}+\Delta}$.
 The function $f(y)$ describes the depinning phenomenon : equal to one when $y<<1$ (strongly pinned DL), it increases rapidly when $y>1$. A simple choice is here $f(y) \sim \exp(y)$. Other choices lead to qualitatively similar results.\\
\indent
It is essential to include the effect of thermal depinning. This can be easily done within the SV regime by taking $L_c(T) \sim L_c(0)\frac{U_0}{T}\exp((\frac{T}{U_0})^3)$ and $\Delta(T) \sim \Delta(0)(\frac{L_c(T)}{L_c(0)})^{-6/5}$ if $T > U_0$ \cite{bible}.
This, together with the above definitions and with the use of renormalized quantities, leads to the self-consistency equation

\begin{equation}
x=\frac{1}{1+ \frac{1}{4c^2_L x} \frac{E^R_{el}}{\nu E^R_{el}+\Delta^R}
 \exp(\frac{T}{\nu E^R_{el}+\Delta^R})\exp(-\alpha\frac{E^R_{el}}{T+\Delta^R})}
\end{equation}

A similar expression could be written in the bundle regime, in which one notices that the plastic susceptibility is larger by a factor $(\frac{\Xi_0}{a_0})^2$ due to weaker DL pinning.
The solution of Eq. (1) is obtained as a function of $\tau = \frac{T}{E_{el}}$ and $\delta = \frac{\Delta(T)}{E_{el}}$ which control the importance of thermal and disorder fluctuations respectively. 
First, in the case of zero disorder, one sees from Eq.(1) that the screening effect becomes dramatic when $T \sim E^R_{el}$, which determines the ``pure'' melting temperature. Here the constants $\nu$ and $\alpha$ are chosen so as to yield the melting transition at $T^0_m = E_{el}$.
 We take $\nu = 0.5$ and $\alpha = 8$. 
 The transverse correlation length $R_1(\tau)$  is plotted on Fig.1.
At the transition this corresponds to a correlation length of order $10^3$ unit cells e.g. a quasi-perfect lattice ($x \sim 0.8$, a value depending on the choice of $\alpha$).\\
\indent
For non-zero disorder we take as a reference the field $B_e(0)$ which marks at $T=0$ the transition to the disordered phase in the Lindemann-based analysis \cite{Giam,Kierf,EN}. Then one defines $b=\frac{B}{B_e} \sim \delta^{5/2}$ (SV regime)
, and similarly $\tau_e = T/T_e=\tau b^{-1/2}$ where $T_e = T_m^0(B_e)$ (see Fig.3). First, for weak disorder, the transition remains first order but occurs at  $T_m < T^0_m$, in qualitative agreement with experiments \cite{melt,Peak} and theory \cite {DF}. 
At $T=0$ the correlation length $R_1$ is still very large.
On the other hand, for strong enough disorder ($\delta > 1.7$), the transition becomes a smooth crossover. This allows to define a critical point $(B_{cr}, T_{cr})$ ending the first-order line (Fig.3). 
Taking for instance $U_0 = 0.8 T_e$ one finds that $T_{cr} \sim 0.52 T_e$ and $B_{cr} \sim 1.65 B_e$ , with $x_{cr} \sim 0.28$. This corresponds to a rather small $R_1 \sim 5 a_0$. The transition can also be represented by plotting $C^R_{66} \sim Bx(B)$ as a function of $B$ at different temperatures (Fig.3).
It is linear at low fields and low temperatures, then bends down and sharply drops at the melting field $B_m(T)$ if $T>T_{cr}$ (first-order transition) or goes smoothly through a maximum if $T<T_{cr}$. 
 Physically, for $T<T_{cr}$ DL's are more strongly pinned thus not mobile enough to drive a sharp collapse of the shear modulus. However, even limited DL fluctuations (or order $a_0$) can have a sizeable screening effect. 
Provided the drop of $C_{66}$ is identified with a melting line, our model is in qualitative agreement with experiments \cite{Peak,Delig}. 
The case of $Bi_2Sr_2CaCu_2O_8$ is special, since experiments provide stronger evidence for a first-order field-induced transition at low $T$. However it might also involve a dimensionnal crossover, which is not present in our 3D model.\\
\indent
Let us now analyse the consequences of the present scenario on vortex dynamics.  First of all, it means that DL's also move to screen elastic stresses, e.g. the dynamics is in reality {\it plastic}. A  refined theory should take a detailed account of the dynamics of both elastic domains and DL's as a function of the length scale $L(J)$ involved at a given current J.
 Here we simply analyze critical currents and  collective creep (CC) within the renormalized elastic picture, and focus on the boundary between SV and 3D behaviours, defined by $L(J) \sim L_0^R$. It depends crucially on the vortex interactions thus on the effective shear modulus.  Here, softening clearly {\it favours} a SV regime by {\it decreasing} the vortex correlations. First, for the SV critical current $J^c_{sv}$, $L(J^c_{sv})=L_c$ and 3D pinning occurs for $Bx(B) > B_{sb}\sim \frac{\phi_0}{2\pi L^2_c}$. Since $Bx(B)$ has a maximum (see Fig.3), two situations are possible, depending on whether $\max(Bx(B)) $ is larger or smaller than $B_{sb}$.
 In the first case, occurring at low temperature, a 3D regime occurs followed by a {\it reentrant} SV regime at high fields (Fig.2). In this case $x(B,T)$ should in principle be recalculated in the 3D regime by generalizing Eq.(1), but this would not change the boundaries of the SV regime (Fig. 3). In the second case, at higher temperatures, the bundle pinning regime is never reached, vortices remain pinned individually. Then one must look for a bundle collective creep (CC) regime, and examine the frontier between SV and bundle CC. 
It occurs at a current density $J_{sb}(B) \sim J^c_{sv}(\frac{B}{B_{sb}})^{7/10}$ \cite{bible}. As before, a reentrant boundary is obtained by using the ``rescaled'' field $Bx(B,T)$ instead of $B$, and is plotted on Figs.2 and 3.
  While a reentrance of SV critical currents was qualitatively suggested in the past as an explanation for the ``peak effect'' \cite{Kes,Vin}, we here underline a similar phenomenon for creep, and provide a quantitative model. It is also much general since for small enough currents, in absence of screening, the  dynamics would always be 3D. This has important consequences on CC behaviour. 
Indeed, in the 3D CC regime, if {\it elastic}, vortex correlations result in a faster divergence of the creep barriers as $J$ goes to zero : while the exponent $\mu$ in $U(J) \sim U_0(\frac{J_0}{J})^{\mu}$ is small, $\mu_{sv} \sim \frac{1}{7}$ in the SV regime, it is larger, $\frac{7}{9} < \mu_b < \frac{5}{2}$ in the elastic 3D regime \cite{bible}.
This ``elastic '' analysis strongly disagrees with experiments performed
 in crystals of $YBa_2Cu_3O_7$ \cite{Abu}, $(Nd,Ce)CuO_4$ \cite{Yesh} and $(Ba,K)BiO_3$ \cite{Klein,Comm} as well as HTSC films \cite{Dek} or multilayers \cite{Trisc}. Crystals show a ``second peak'', together with {\it non-monotonous} creep exponents, all features incompatible with the elastic analysis. Due to creep, the peak cannot be strictly attributed to a non-monotonous critical current, and a crossover in creep regime is possible \cite{Kru}. 
However, this last scenario assumes bundle CC above the peak, with large $\mu_b$, while it was shown \cite{Klein,Comm} that the CC scaling still holds provided one uses a field-dependent $\mu(B)$ which {\it decreases} at high fields. 
On the other hand, the reentrance found in our model nicely explains these results : as the field increases, the exponent $\mu$ has a maximum inside the bundle CC region and eventually goes back to the SV value as the SV regime is recovered. We thus claim that such a non-monotonous $\mu(B)$ is a characteristic feature of ``plastic'' collective creep. 
Remarkably, in $(Ba,K)BiO_3$ \cite {Klein}, the dynamics above the second peak cross over to a VG transition, the dynamical exponent reaching a value close to the SV exponent $\mu_{sv} = \frac{1}{7}$. 
Such small values seem ubiquitous close to VG transitions, an appear as a signature of a {\it renormalized SV creep regime}. We thus show that softening is an essential phenomenon to account for the dynamics at and above the peak. 
More generally, the present scenario could also be applied to the critical current ``peak effect'' \cite{Kes,D'Anna,Bhat,Vin} occurring close to $B_{c2}$, provided the full $B-$dependence of $C_{66}$ is used. Recent neutron scattering experiments in $Nb$ indeed show the succession of vortex lattice, hexatic and amorphous solid as $B$ increases towards $B_{c2}$ \cite{Gam}.\\
\indent
The physical picture in the reentrant SV regime is the following : each vortex line wanders laterally on distances much larger than $a_0$, thanks to edge DL's. Softening make the effective correlation between vortex lines  much smaller, providing a simple mean-field picture of an entangled hexatic solid. 
However, it does not allow to fully describe the highly disordered phase. First, above the first-order drop ($T>T_{cr}$), $C_{66}^R$ is not exactly zero therefore the phase from this point of view is still a solid. 
Similarly, at low temperatures and high field the shear modulus does not vanish, as expected close to a VG transition. In fact, other topological defects such as vacancies or intersticials should be considered \cite{Vac,Yu}, also a more powerful approach is necessary to describe a truely amorphous glassy phase
 with possibly second-order (solid-solid and solid-liquid) transitions, joining at the critical point ($B_{cr},T_{cr}$). Our approach, limited to screening induced by short-range fluctuations, should be improved to describe such transitions. One can nevertheless try to extend the present screening idea to the VG transition proposed by Fisher \cite{vg}. 
 First, in an elastic picture, thermal depinning is characterized by the equality between $T$ and the pinning energy 
 gained at the scale of the temperature-dependent Larkin correlation length $L_c(T)$. 
This length becomes very large at high temperature where disorder is effectively washed out. On the other hand, in the vortex glass transition picture, a glass correlation length
$\xi_{vg}(T)$ is defined \cite{vg}. The behaviour at length scales $L < \xi_{vg}$ is critical, e.g. characterized by the equality between $T$ and the vortex glass energy fluctuations. 
 In the plastic vortex lattice picture, these fluctuations involve the pinning energy of DL's. When the DL density is so large that the vortex solid becomes amorphous, due to flux cutting and other defects, a relevant length scale ${\cal L}_c(T)$ must be defined. 
In this phase the effective disorder should be renormalized down by large scale fluctuations. We suggest that through this phase ${\cal L}_c$ go smoothly towards the vortex glass correlation length $\xi_{vg}(T)$, with $\xi_{vg}(T) \sim \xi_0 (T_{vg}-T)^{-\nu}$ \cite{vg}.
The above screening model can be simply modified by replacing $\Xi_0$ by $\xi_{vg}(T)$. Being critical, at the length $\xi_{vg}$, $T$ is of the order of the pinning energy therefore $<u^2_d> \sim \xi^2_{vg}$ in the expression for the plastic susceptibility $\chi$. This, together with $E^R_{el} \sim \Delta^R \sim T$ leads to $x^{-1} \sim 1 + \frac{1}{4c^2_L x} (\frac{\xi_{vg}}{a_0})^2 $
which close to the VG transition amounts to $C^R_{66} \sim (T_{vg}-T)^{2\nu}$. Full softening is completed at the vortex glass transition where $C_{66}^R$ becomes zero and the vortex glass melts.\\
\indent
In conclusion, we have proposed a model for the softening and melting of a vortex lattice in presence of point disorder. We find a critical point ending the first-order line, and a continuous crossover towards an hexatic glassy phase at low temperature, associated with a maximum in the shear modulus. 
Softening of the vortex solid manifests itself in vortex decorrelation thus a reentrance of single vortex pinning and creep,  providing an explanation for the anomalous creep exponents. 
This model allows to test quantitatively the possibility of disorder-induced plasticity, however much more work is required to analyze in detail the plastic dynamics and to properly describe the (amorphous) vortex glass transitions.\\
\indent
The author acknowledges hospitality from Dipartimento di Fisica, Universita La Sapienza, Roma  and support from Istituto Nazionale di Fisica della Materia. 
He is grateful to C. Dasgupta for helpful discussions, held thanks to a PICS (CNRS) project with JNC, Bangalore, as to T.Klein and F. Thalmann in LEPES.

\end{multicols}
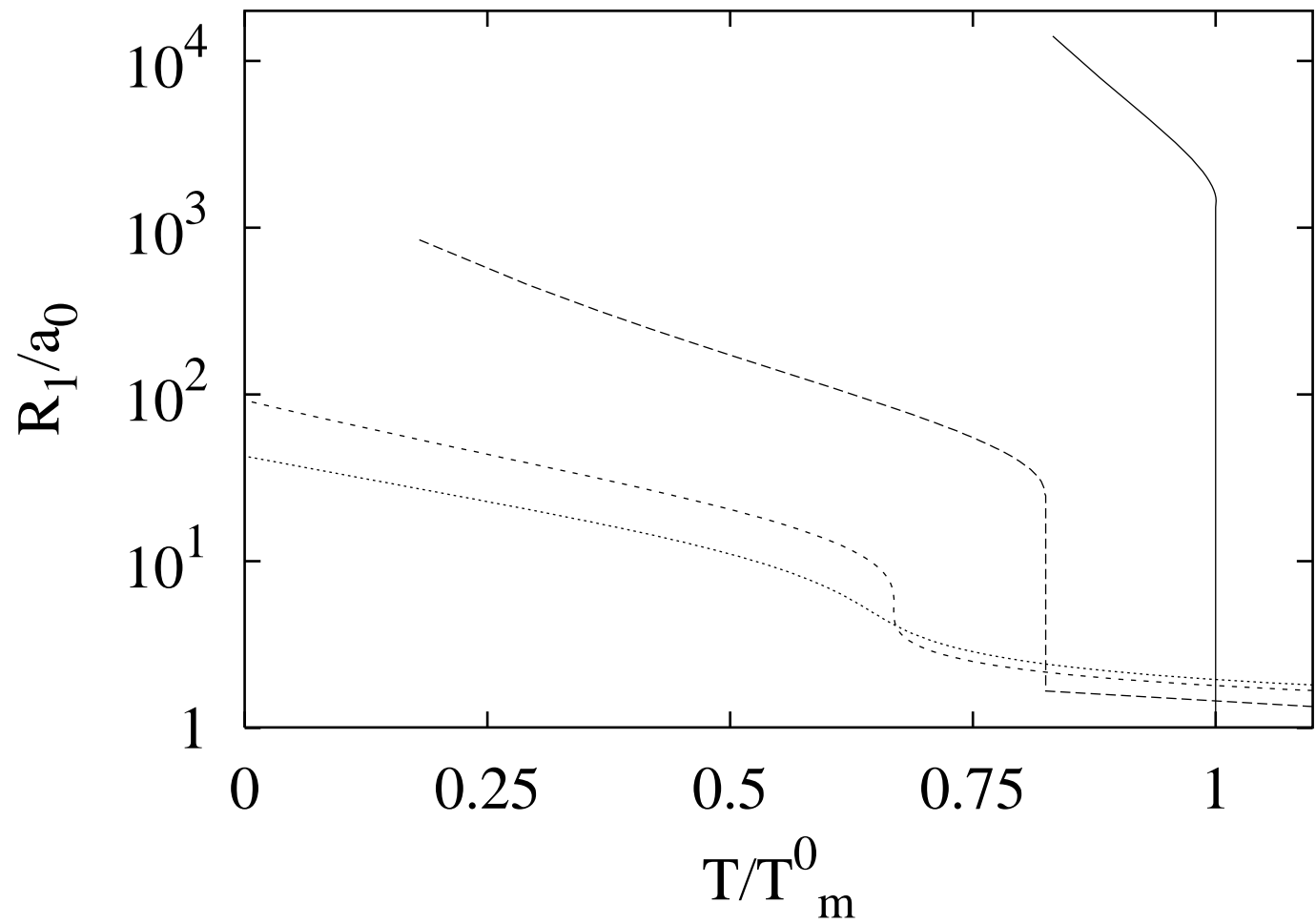
\begin{figure}
\caption{Transverse correlation length $\frac{R_1}{a_0}$, as a function of 
$\tau=\frac{T}{T_m^0}$ and increasing disorder, from top to bottom : 
$\delta=0, 1, 1.71$ (critical),$2$.}
\end{figure} 

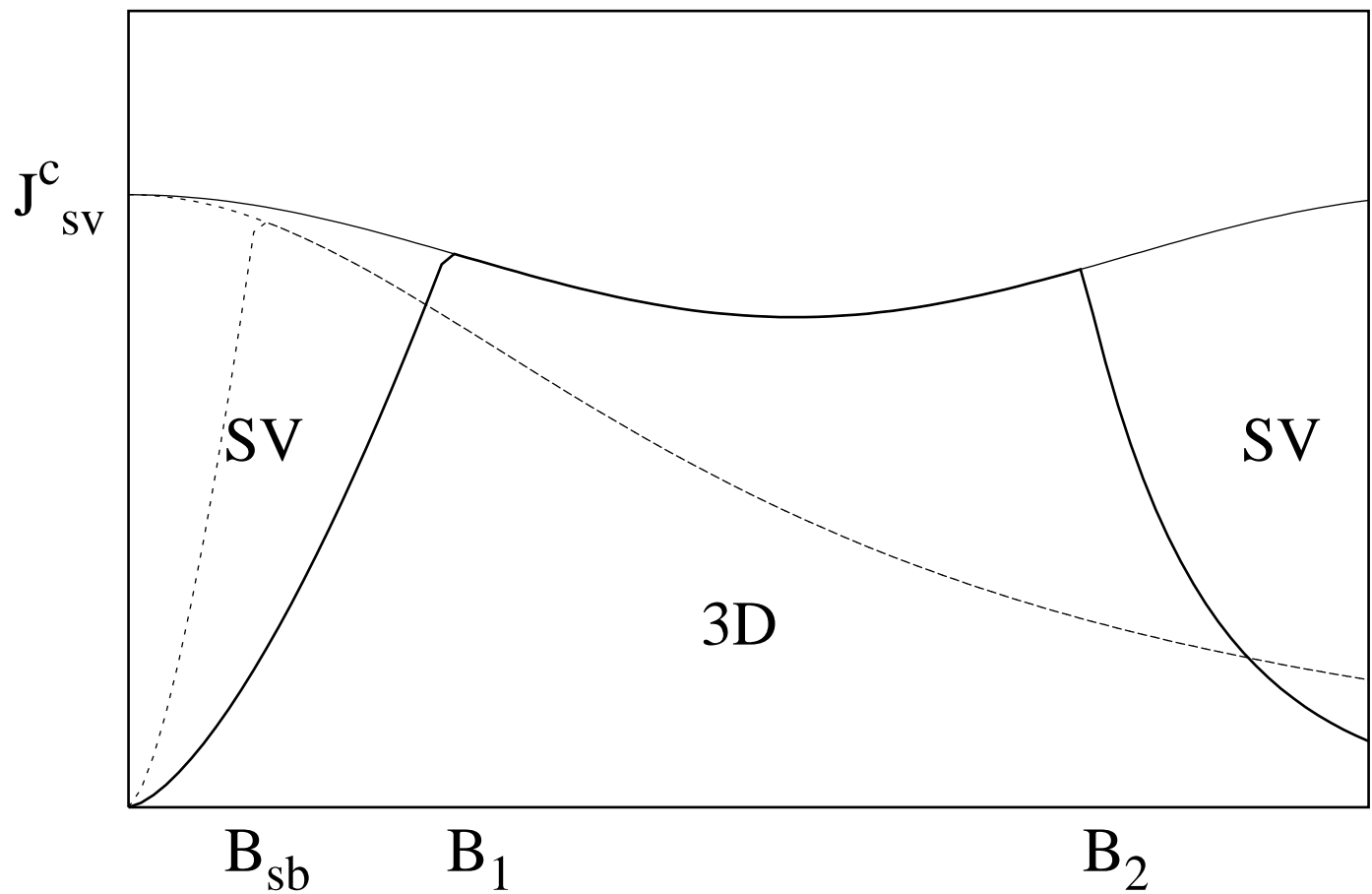
\begin{figure}
\caption{Schematic $(B,J)$ phase diagram showing the SV and 3D creep regime boundaries with renormalized $C_{66}$ (thick lines) and the reentrance at high fields. $B_1$ and $B_2$ are the limits of SV critical currents. The dotted lines stand for the pure elastic analysis.}
\end{figure}

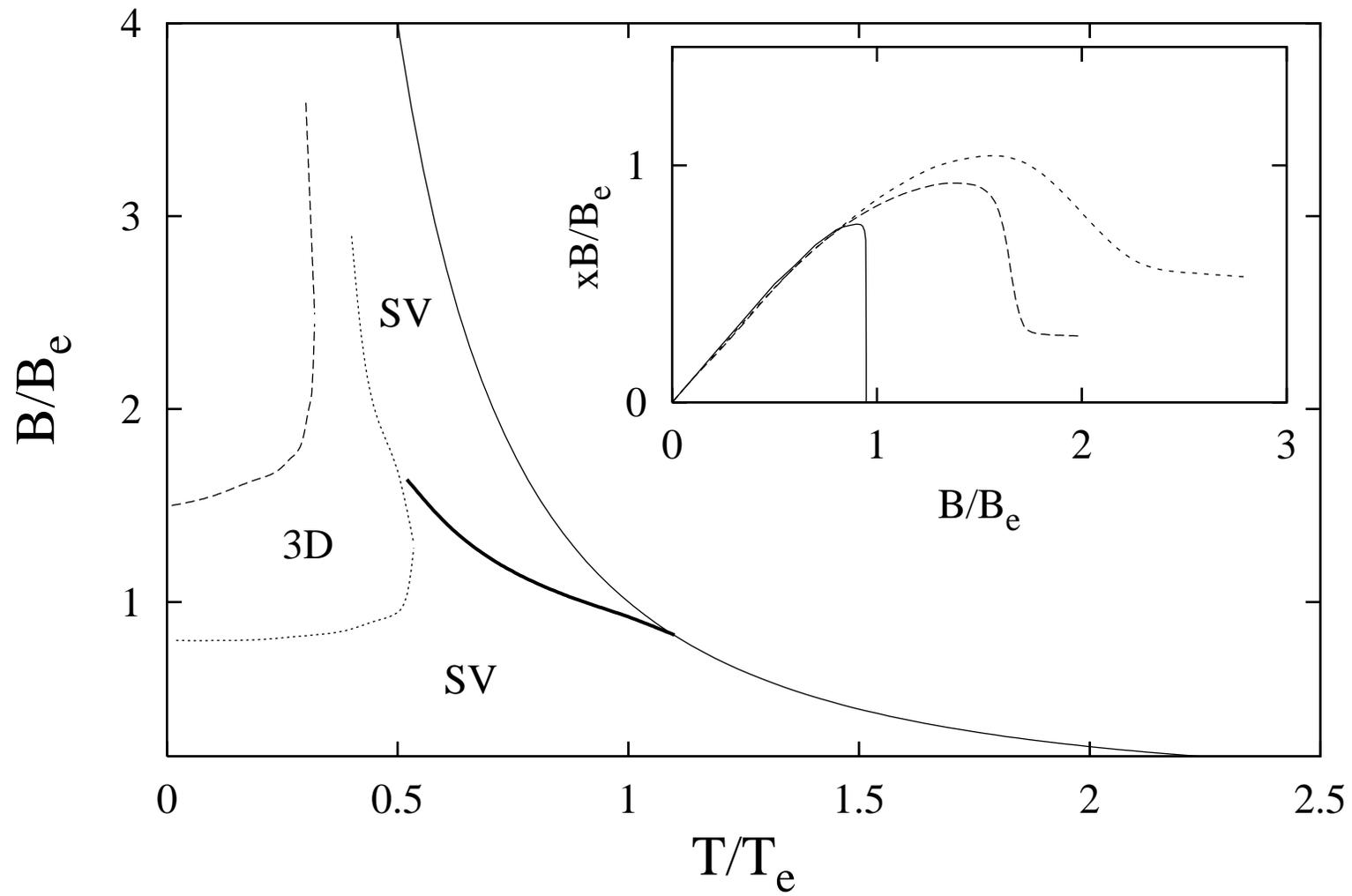
\begin{figure}
\caption{Phase diagram showing the pure melting line
 (thin line), the first-order melting with disorder
 (thick line) and the boundaries of SV pinning (dashed,
 $B_{sb}=1.5 B_e$) and SV creep (dotted, $J=J_{sb}(0.8B_e)$);
 $\frac{U_0}{T_e} = 0.8$ (see text). Inset shows the renormalized
 shear modulus in units of $B_e$, for increasing temperatures :
from top to bottom, $\frac{T}{T_e} = 0.45, 0.52$ (critical), $1$.}
\end{figure} 

\end{document}